\documentclass[12pt,preprint]{aastex}
\def\beq{\begin{equation}}
\def\eeq{\end{equation}}
\def\bey{\begin{eqnarray}}
\def\eey{\end{eqnarray}}
\def\pc{\, {\rm pc} }
\def\kpc{\, {\rm kpc} }
\def\msun{M_\odot}

\def\lsim{\mathrel{\raise.3ex\hbox{$<$\kern-.75em\lower1ex\hbox{$\sim$}}}}
\def\gsim{\mathrel{\raise.3ex\hbox{$  $\kern-.75em\lower1ex\hbox{$\sim$}}}}

\def\kms{\, {\rm km \, s}^{-1} }

\begin{document}

\title{Tidal Disruption of The First Dark Microhalos}
\author{
HongSheng Zhao\footnote{National Astronomical Observatories, Chinese Academy of
Sciences, Beijing 100012, PRC} $^{ ,}$\footnote{SUPA, School of Physics and Astronomy,
University of St Andrews, KY16 9SS, UK} , 
Dan Hooper\footnote{Fermi National Accelerator Laboratory, Batavia, IL  60510-0500} , 
Garry W. Angus$^{2,1}$, 
James E.~Taylor\footnote{Astronomy Department, Caltech, mc 105-24, 1200 East
California Blvd., Pasadena CA 91125},
Joseph Silk\footnote{Astrophysics, University of Oxford, Denys Wilkinson Building, Keble Road, Oxford OX1 3RH, UK}
}
%\date{\today}

\begin{abstract}
We point out that the usual self-similarity in cold dark matter models is broken by 
encounters with individual normal galactic stars on sub-pc scale.  Tidal heating and stripping must have redefined the density and velocity structures of the population of
the Earth-mass dark matter halos, which are likely to have been the first bound structures to form 
in the Universe.  
The disruption rate depends strongly on {\it galaxy types} and the orbital distribution of the microhalos; in the Milky Way, stochastic radial orbits are destroyed first by stars in the triaxial bulge, microhalos on non-planar retrograde orbits
with large pericenters and/or apocenters survive the longest.  
The final microhalo distribution in the {\it solar neighborhood} 
is better described as a superposition of filamentry microstreams rather 
than as a set of discrete spherical clumps in an otherwise homogeneous medium. This has important consequences to our detections of microhalos by direct recoil signal and indirect annihilation signal. 

\end{abstract}
%\pacs{95.35.+d; 98.35.Gi; 98.35.Jk}
\maketitle

\section{Introduction}

Cold dark matter (CDM) has had considerable success in accounting for the observed 
large-scale structure of our Universe. On galactic and sub-galactic scales, however, 
the abundance of low-mass structure and the degree of concentration of dark matter halos 
predicted by CDM simulations have provoked a considerable amount of discussion. 
To clarify how CDM behaves on the smallest scales, Diemand and 
collaborators recently performed the highest resolution numerical simulations 
of dark matter clustering to date (Diemand et al. 2005). These simulations start with an initial spectrum 
of density fluctuations extending down to the free-streaming mass of the dark matter 
candidate, which for a generic weakly interacting 100 GeV particle (such as a supersymmetric 
neutralino) is roughly $\sim 10^{-6}\,(t_{fo}/10^{-8} {\rm s})^{7/4}\, M_\odot$, where freeze-out 
occurs at an epoch $t_{fo} \sim 10^{-8}\,(m_\chi/100\, \rm GeV)\rm \,s$. 
As expected from scaling the results of simulations on larger scales, fluctuations
in this initial distribution collapse and virialize to form structures with roughly 200 
times the background density at a redshift of $z\sim 50.$ This first generation of 
'microhalos' survives to some degree as substructure in the larger halos that form 
subsequently in the simulations. 
Extrapolating to the present-day, Diemand et al. 2005 suggest that $10^{15}$ Earth-mass 
clumps should survive in the halo of the Milky-Way, amounting to about 0.1\% of its 
total mass. If this were the case, the nearest Earth-mass clump would be on the order 
of 0.1 pc distant from the Earth and would have a typical size of 0.01 pc. 
Such nearby clumps might be observable as sources of gamma-rays, produced in dark matter 
annihilations. They might also contribute to diffuse cosmic ray fluxes of positrons or 
anti-protons. In order to motivate experimental searches, however, it is important to 
determine whether the predicted small clumps of dark matter  
would actually survive the grainy tidal field due to stars while crossing the Galactic bulge 
and disk $\sim 100$ times.

Here we estimate the rate at which Earth-mass clumps are disrupted 
in stellar encounters. 
We show that the nearest clumps in a galaxy like the Milky Way are likely to be tidally 
disrupted by repeated encounters over a Hubble time. Nevertheless, tidal debris from 
individual clumps may produce distinct microstreams that are potentially observable in 
dark matter detection experiments.

\section{A Semi-analytic Mass-loss Estimate}
 
Extended bodies generally have a complex response to external tidal heating (Gnedin \& Ostriker 1999). 
We can estimate the net effect of this process by applying the basic scaling 
from the semi-analytic mass-loss model of Taylor \& Babul 
(2001), which has been shown to match tidal heating and mass loss 
rates in high-resolution numerical simulations. According to this model, 
whenever a system on a general orbit spends time $\Delta t$ in a strong tidal field, 
that is a field in which the instantaneous tidal limit for the object $r_{\rm t}(t)$  
(calculated as in the circular-orbit case -- cf. Binney \& Tremaine 1987) 
is smaller than its size $r$, it loses a fraction $\Delta t/ t_{\rm orb}$
of the mass outside $r_{\rm t}(t)$, where $t_{\rm orb} = 2\pi r(t)/v(t)$ 
is the instantaneous orbital period. 
%We note that this scaling applies to relatively slow encounters (where the duration of the encounter is longer than or comparable to the internal dynamical time of the system). Rapid encounters produce even more heating and mass-loss, but the scaling is more complicated (\cite{gnedin,taylorb}). In what follows we will only consider the basic scaling, which should provide a conservative lower limit on the true mass loss rates.

To estimate when material at radius $r$ in a microhalo will experience strong shocks, 
we need to calculate the minimum impact parameter such that $r_{\rm t}(t) < r$.
This is equivalent to the condition that the tidal force generated during an 
encounter with a star of mass $m_*$ at an impact parameter $b_{sh}$, 
${2 G m_* x / b_{sh}^3}$, exceeds the restoration force for a
small displacement $x$, ${4\pi G \rho_{<r} \over 3} x$,
 where $\rho_{<r}$ is the
mean density of the microhalo interior to $r$.
For a typical microhalo in the simulations by Diemand et al. (2005), the density at
the half-mass radius is $\sim 1\msun\pc^{-3}$, while the highest resolved density (at ~0.1
$r_{200}$) is $\sim 10$ times higher. We do not expect a divergent (phase space)
density in these smallest halos.  Even if a cusp existed below the simulation
resolution, it would contain less than 3\% of the microhalo's mass,
extrapolating from the profile given by Diemand et al. (2005).
Thus for a rapid encounter with a solar-mass star, impact parameters of 
$b_{sh} \sim \left({m_* / \rho_{\rm mic}}\right)^{1/3} \sim 0.5\pc$ or 
less will produce a strong shock over a range of radii containing 97\%  or more 
of the mass in the microhalo. $\rho_{\rm mic}$ is the average density within 10\% of the virial radius $R_{\rm vir}$ and can be thought of as a constant until the microhalo is virtually destroyed as shown in Angus \& Zhao (2006, hereafter AZ, figures 7c and 12b).

A microhalo moving at a relative speed of $V_r \sim 300\kms$
through a star field with a number density $n_*={\rho_* / m_*}$ 
will encounter stars with an impact parameter $\lsim b_{sh}$
at a rate of $\dot{N}_{\rm enc}=n_* V_r \pi b_{sh}^2 \sim {n_* / 0.004 \pc^{-3}}$ 
per million years.  Each encounter shocks the microhalo for a time 
$\Delta t = 2b/V_r \sim 3300\,$ years, and liberates a fraction of the microhalo's 
total mass. This fraction, from the scaling in Taylor and Babul (2001), is 
roughly  ${\Delta m / m} = \Delta t/t_{\rm orb} = {v_{\rm orb} / \pi V_r} \sim 10^{-4}$. 
Here, $t_{\rm orb} \equiv {2 \pi b_{sh} / v_{\rm orb}} \sim 30\,$ Myr is the period 
of a circular orbit of speed $v_{\rm orb}=\sqrt{G m_* / b_{sh}} \sim 90\,$m/s 
around the solar-mass perturber at a distance of $b_{sh} \sim 0.5\pc$. 
Thus, the mass of a microhalo decreases exponentially, with the e-folding time given by 
${1 / \tau} \equiv - m^{-1}{\rm d}m(t)/{\rm d}t = \dot{N}_{\rm enc} {\Delta m / m} 
= G n_* m_* (G \rho_{\rm mic})^{-1/2}$,
where we have expressed $b$ in terms of the parameters of the satellite.  
Averaged over a Hubble time, we find that a microhalo's mass decays exponentially,
$m(t)=m_0\exp(-t/\tau)$, where the e-folding time and the present mass are given by
\beq\label{tau}
\ln {m_0 \over m_{\rm now}} = {10 {\rm Gyr} \over \tau } \approx  
\left( {\left<\rho_*\right>  \over 0.004 \msun\pc^{-3} } \right) 
\left( {\rho_{\rm mic} \over 10\msun\pc^{-3} } \right)^{-1/2},
\eeq
where $\left<\rho_*\right>$ is the mass density of 
galactic (disk, bulge and halo) stars averaged along the orbit 
of a microhalo over the past 10 Gyrs (at earlier times, few stars might have formed). 
In general, $\left<\rho_*\right>$ is a function of the 
pericenter, apocenter and vertical height of the microhalo orbit.  E.g., a microhalo on a thick 
disk orbit of height 1000 pc near the solar neighborhood
would have made $\sim 100$ disk crossings and 
gone through star fields of an accumulated column density $\sim 10^4\pc^{-2}$ by today. 
It goes through the galactic disk with a mean surface density of 46\,$M_{\odot}$pc$^{-2}$ or 
an average volume density $46 M_{\odot}$pc$^{-2} / 2000\pc \sim 0.023\msun\pc^{-3}$, 
thus this microhalo's mass is predicted to have decayed by about 6 e-foldings.

\begin{figure}
\plotone{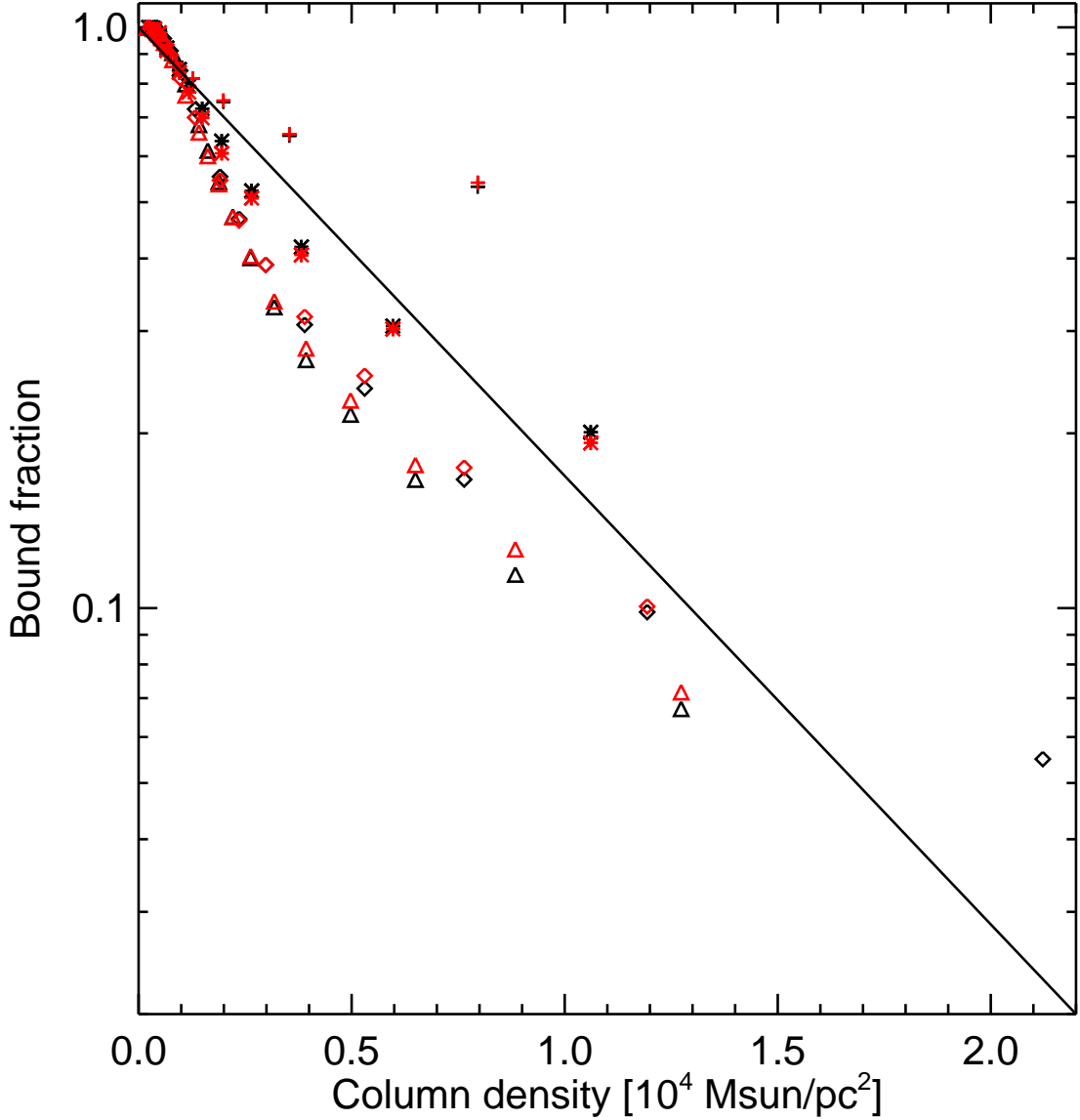}
\caption{shows the post-impulse fraction of bound microhalo particles vs.   
the equivalent column density $\Sigma_*=M_*/(\pi b^2)$ of the perturbers, where 
symbols denote results from a set of numerical experiments 
of stellar encounters with microhalos of initially tangential anisotropy (TA, red symbols) nd radial anisotropy (RA, black symbols).  In these experiments
a perturber with $M_*=0.1\msun$ (plus), $0.3\msun$ (asterix), $0.6\msun$ (diamoNds)
and $1\msun$ (triangles) passes a $10^{-6}\msun$ microhalo with relative velocity $V_r=200\kms$ and different values of the impact parameter $b$ ranging from penetrative ($b<R_{vir}$ to non-penetrative encounters ($b>R_{vir}$). 
For the same mass column density, many low-mass penetrative perturbers are not as efficient
as few high-mass non-penetrative perturbers in terms of stripping the microhalo;  
stripping is unimportant for sub-stellar perturbers in general.    
Our semi-analytical prediction assuming $\rho_{\rm mic}=10\msun\pc^{-3}$ is represented as the solid line.}
\label{angus2}
\end{figure}
Another way to think of the semi-analytical disruption rate is to note that the logarithm 
of the bound fraction 
$\ln {m_{\rm now} \over m_0} \propto \int \rho_* dt \propto {\Sigma_* \over V_r}$,
where $\Sigma_*$ is the stellar mass density encountered.  
To get an indication of how robust the semi-analytical massloss rate is, 
we have also run N-body simulations of a simplistic 
single encounter between a star and a microhalo for comparison.
We use the tree-code of Vine and Sigurdsson 1998.
We use a total particle
number of $10^5$ and softening of $10^{-4}$pc consistent with interparticle
separations of 3.5$\times 10^{-4}$pc.  The tree code tolerance
parameter, $\theta$ is the cell size divided by particle separation and is set to
0.6 in these simulations. 
Initially the microhalo particles were distributed in a spherical NFW density profile
with a $R_{\rm vir}=1.6 R_c =0.01$pc and a total mass of $10^{-6}\msun$. Two sets of velocity space initial conditions were used; tangentially anisotropic (TA) microhalos with randomly inclined circular orbits $\bar{V_r^2}(r)=0$ and radially anisotropic (RA) microhalos with $\bar{V_r^2}(r)={\bar{V^2}(r) \over 2} ={1 \over r \rho(r)} \int_r^{R_{vir}} V_{\rm c}^2 \rho dr$, where $\rho(r)$ and $V_{\rm c}$ are the microhalo's internal density and circular velocity respectively (cf. Angus and Zhao 2006).  Two-body relaxation time, measured by the mean squared normalised
energy drift after $10^4$ time steps using $10^5$ particles for an isolated microhalo, 
is about 3Gyr.  So a microhalo resolved with $10^5$ particles should be free from 
numerical relaxation on time scales of 15Myr, i.e., the dynamical time scale of the microhalo.  
To measure the effect of encounters with stars, we shoot a single star of mass 
$M_*=0.1,0.3,0.6,1\msun$ with a relative velocity $V_r=200\kms$ towards a microhalo 
with a closest approach $b=(0.2-1)\times R_{\rm vir}$ for penetrative encounters
And $b=(1-4)\times R_{\rm vir}$ for non-penetrative encounters.
Such an impact is typical for a microhalo passing through a random star field with 
an equivalent star column density $\Sigma_*=M_*/(\pi b^2)$.
The duration of a star passage, say, along the x-axis from $x=10R_{\rm vir}$ to $x=-10R_{\rm vir}$ 
is about $0.0001$ dynamical time of the microhalo, hence microhalo particles are virtually stationary
(moving less than 0.01 percent of its orbital radius) 
during a star passage, so the perturbations are indeed in the impulse regime.   
It appears that our semi-analytical model captures the essential feature in numerical 
simulations of the impulse on $10^{-6}\msun$ microhalos 
if assuming $\rho_{\rm mic} \sim 1\msun\pc^{-3}$ (cf. Fig.1).
However, there are microhalos of a few times Earth mass, and these
can be, e.g., 5 times more massive and denser than the $10^{-6}\msun$ microhalos 
experimented here (cf. Fig.2 of Diemand et al. 2005).  Allowing for more robust microhalos
and to be conservative in estimating the disruption rate of microhalos,
we shall adopt the rate from the semi-analytical model assuming 
$\rho_{\rm mic} \sim 10\msun\pc^{-3}$ hereafter.
Note that the scaling of our semi-analytical model Eq.~\ref{tau} 
also agrees with the literature on impulse-induced massloss in the context of diffuse star clusters, 
e.g., Moore (1996), and Eqs.~6-8 of Wielen (1985) or Eq.~8 of Wasserman \& Salpeter (1994).  

\begin{figure}
\plottwo{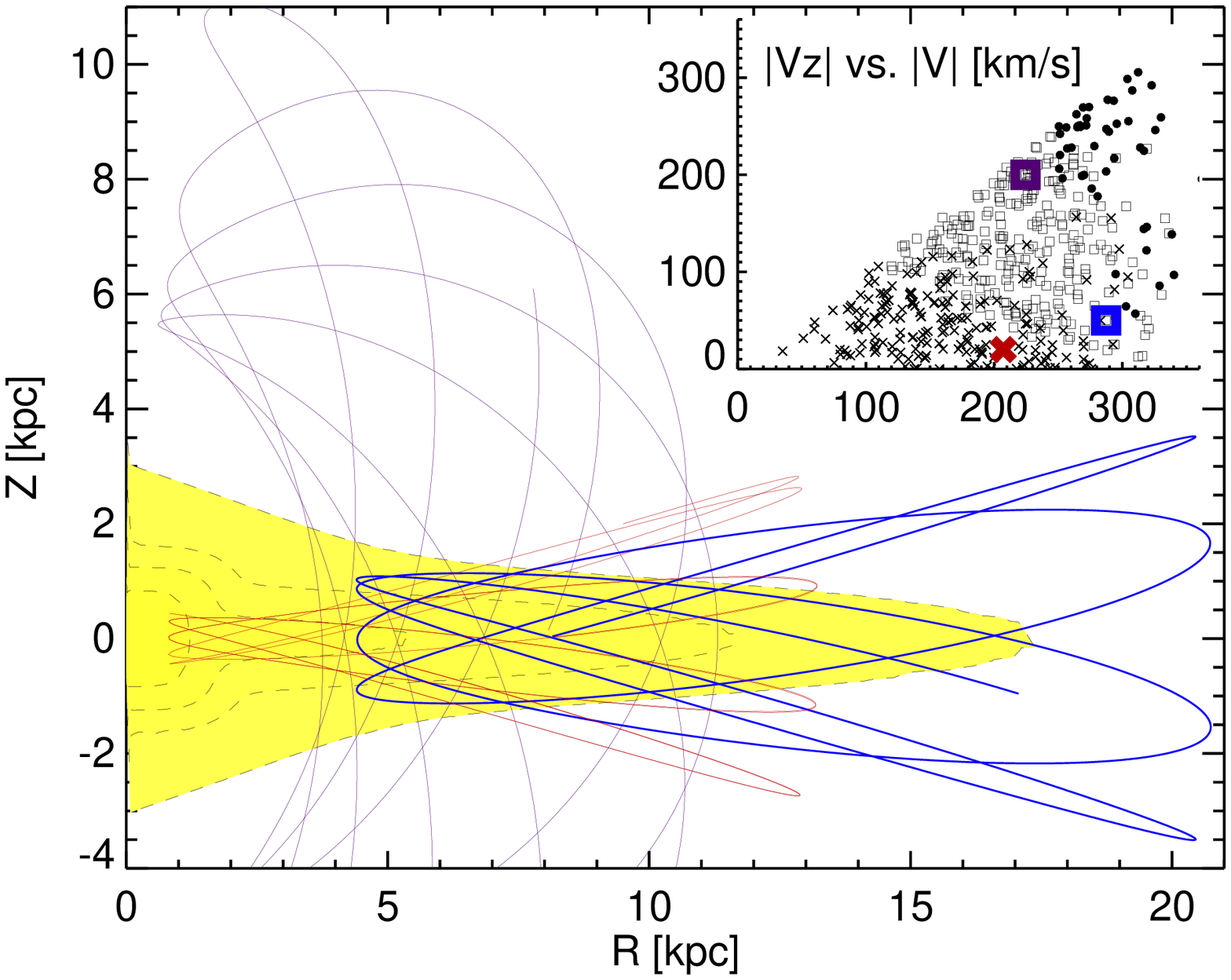}{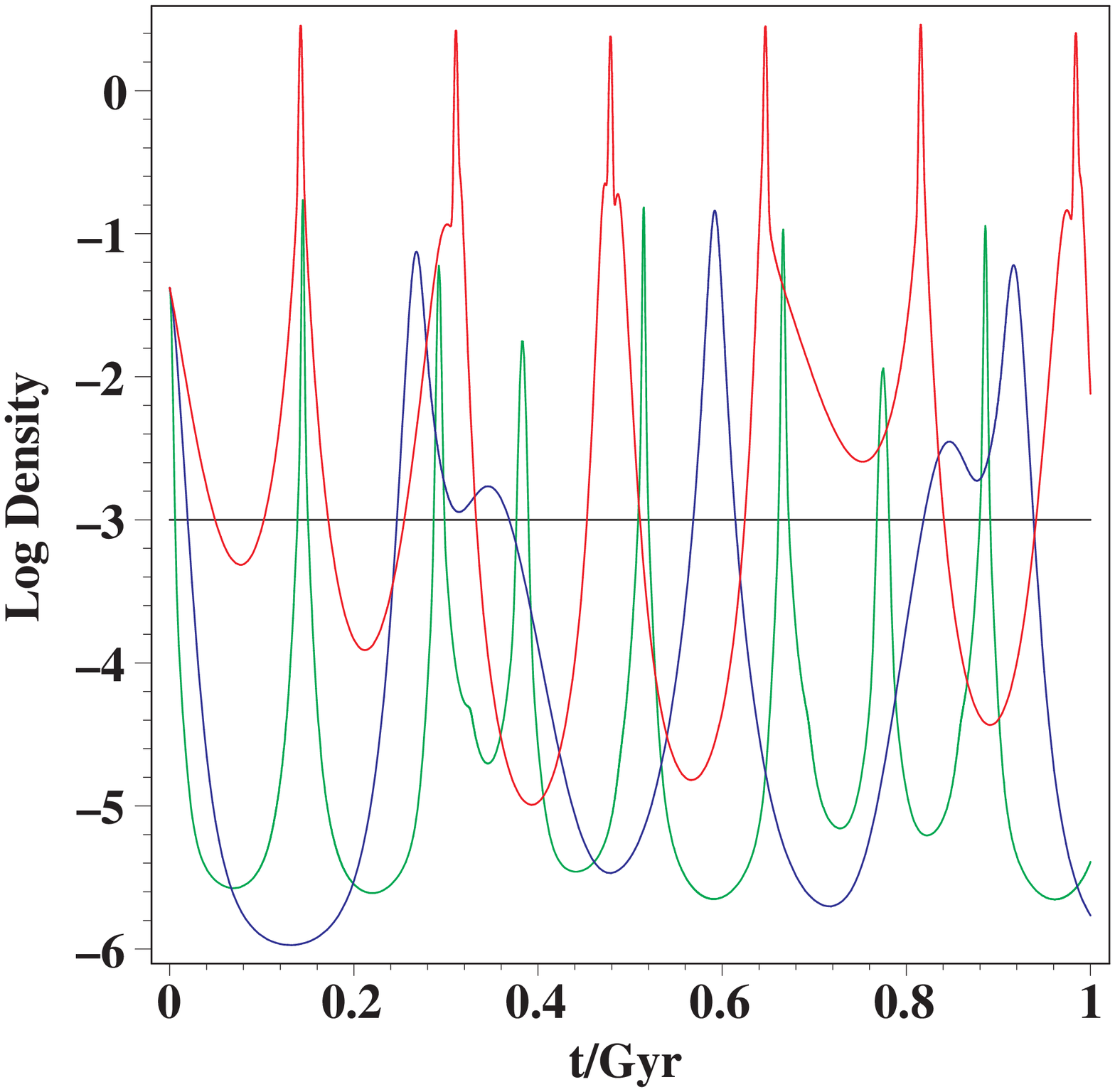}
%\vskip -2cm
\caption{{\it Left panel}: 
The inset shows the scatter diagram of the velocity space (at launch)
for the 500 microhalos launched from the solar circle.  
Microhalos are given different symbols depending on the time-averaged
stellar density along the orbits ($\left<\rho_*\right>\, >0.01\msun\pc^{-3}$ for crosses and
$<0.001\msun\pc^{-3}$ for solid circles and in between for squares).
Note microhalos with low orbital energy and angular momentum are first
to be disrupted. Three typical orbits (larger symbols) are picked out
and shown in meridional plane $Z$ vs. $R$.
The yellow shaded area is a cut of the Besancon star count model of
the Galaxy (Robin et al. 2003)
with equal star density of $10^{-i}\msun\pc^{-3}$ for $i=0,1,2,3$
plotted in contours.
%}\label{angus1}
%\end{figure}
%\begin{figure}
%\plotone{f3.eps}
%\resizebox{7cm}{!}{\includegraphics{f3.eps}}
%\caption{
{\it Right Panel:} shows the star density in $\msun\pc^{-3}$ 
along the three colored orbits as function of time.  
}\label{angus1}\label{oldfig}
\end{figure}

\section{Microhalos on Near-Earth Orbits}

To study the dependence of mass loss on orbital parameters, we have integrated 
orbits of microhalos in a flattened, axisymmetric galaxy potential with a nearly
flat rotation curve:
$
\Phi(R,z)=(220\kms)^2 \ln {\sqrt{R^2+z^2/0.8} \over \sqrt{R^2+z^2}+60\kpc}.
$
We launch 500 orbits from the solar neighborhood, $(R,z)=(8\kpc,0)$, with a Gaussian 
velocity distribution ($150\kms$ dispersion in each direction).  We read off the 
stellar density along the orbit using the Besancon model of the Galaxy 
(Robin et al. 2003), and then take a time average of the star density. The Besancon model is a widely accepted detailed prescription of the number density of stars and stellar remnants of different ages and masses in the Galactic spheroids (bulge and halo) and disks (thin and thick).  
A few examples of the microhalo orbits are given together with 
the star count model in Fig.~\ref{angus1}.  These simulations suggest that the population of microhalos 
with apocenters less than 12 kpc has $\left<\rho_* \right> \sim 0.002\msun\pc^{-3}$, 
and are hence past their half-life (purple orbit in Fig.~\ref{angus1}). The more planar orbits, 
with $|Z| \lsim 4(R/20)\kpc$ and apocenter $R \lsim 20\kpc$ (blue orbit in Fig.~\ref{angus1}), 
have decayed by approximately 1 e-fold with $\left<\rho_*\right> \sim 0.004\msun\pc^{-3}$. 
Complete destruction should happen to disky orbits penetrating into the bulge; e.g., the 
red orbit in Fig.~\ref{angus1} goes through a typical column density of $500\msun\pc^{-2}$ with 
each disk crossing at a small angle, and has on average
$\left<\rho_{*} \right>$ $\gsim 0.075\msun\pc^{-3}$, corresponding to 
about 19 e-folds.  
In general, we observe a strong correlation between the orbital shape 
and the disruption rate.  
Orbits which go through dense regions of the Galaxy  are likely disrupted due to tides
(cf. Fig.~\ref{oldfig}).  

Also plotted in the inset of Fig.~\ref{angus1} is the zone of destruction in the velocity space
using 500 microhalos launched from the solar neighborhood. There is a
clear trend for halos with more planar (smaller $|V_z|$) orbits, especially those co-rotating with the disk,
and orbits with lower energy (hence smaller apocenters) to enter regions of higher density.
The sample averaged density is $0.008\msun\pc^{-3}$ or $0.02\msun\pc^{-3}$ for median or mean density, so microhalos on Earth-crossing 
orbits are likely to have been severely stripped. About 10\% of the microhalos are in 
low-density regions (below $0.001\msun\pc^{-3}$), and might survive more-or-less intact. 
The chance of survival, however, is likely even more severely reduced in the triaxial, barred and 
evolving potential of the Milky Way, where most orbits are stochastic box orbits which 
pass through the dense center of the Milky Way at some point over a Hubble time. Additionally, it has been shown in AZ that disk shocking and tidal stripping may combine with stellar encounters to destroy microhalos even more efficiently than shown here. 

Of course, these orbits are mearly a sample of all orbits passing the sun.
In fact, the fraction of halos disrupted would be much lower
for high-energy orbits beyond the solar neighbourhood as the tides at larger pericentre are weaker along with the rapidly decreasing fraction of stars. What was also keenly noted by Berezinsky et al. (2005) was that apocentric disk crossings beyond the extent of the thin disk are likely to boost the lifetime of microhalos. The destruction rate is certainly much higher for
low-energy orbits contained inside the bulge or those that have pericentres within the solar radius. These michohalos would be exposed to stronger tides and higher stellar densities added to the fact they will cross the disk twice every orbit. However, these orbits are irrelavent
for direct experiments on Earth, and less important for indirect experiments
because microhalos are barely identifiable from the background
beyond about 0.1pc - 1pc distance from us.

Additionally, the calculations done by Diemand et al. (2005) did not find
self-consistent orbits and their comments on the
current situation in the Milky Way are major extrapolations from
dark-matter only simulations that stop at redshift 26, and only
ever simulate a much smaller total volume. Since halo orbits in
the MW would be strongly affected by the formation of the galaxy. So what we have here is desruption as a function of orbital parameters.

We note that the long term fate of the cusp at the very center of each microhalo is unclear. 
Simulations have shown that systems with a universal density profile may lose
all but 0.1--1\% of their mass and still retain a bound central region Hayashi et al. (2003). 
Such objects may indeed survive to the present-day in the solar neighborhood, but they 
will have little effect on dark matter detection experiments. The tidal debris stripped 
out of microhalos may actually be of greater interest, as discussed below. Nevertheless, AZ (figures 7c and 12b) showed that the inner density of a microhalo ($\rho_{\rm mic}$) is largely unaffected until it is stripped down to the inner shell.

\section{Implications For Direct and Indirect Detection} 

The existence of dark
substructure in our galaxy's halo has important implications for
the prospects of both direct and indirect dark matter detection (Green et al. 2005; Berezinsky et al. 2003; Pieri et al. 2005; Koushiappas et al. 2004).
The presence of substructure affects direct detection rates in a straight forward way: 
If the solar system happens to be located
inside of an overdense region of dark matter, then the rate
observed will be enhanced proportionally to
the density of the clump. If, as is much more likely, our solar
system is not inside a clump, then the rate will be
modestly reduced by the fact that some fraction of the overall
density is contained in substructures and thus does not contribute
to the density of the smooth component.

Zhao and Silk 2005 estimated
that the nearest minihalo of $10^6\msun$ is 1 kpc away with a size of 50pc.
Diemand et al. 2005 estimated that unstripped Earth-mass halos have a filling
factor of $0.5\%$.  Hence in either case it is {\it unlikely} for the Sun to be inside a substructure.  

Tidal debris from encounters with stars changes the
picture substantially.  While the probability to find 
our solar system inside of a $10^{-4}\pc$-sized bound remnant of a microhalo
is on the order of $10^{-8}$, the probability is of order unity
for the solar system to be moving presently inside an (unbound) substructure.
Disrupted streams are inhomogeneous regions with at least 10 times the
volume filling factor of the original unstripped clump.  This is because
at each disk crossing a burst of tidal debris is released due to
encounters with disk stars, and the tails
would lead/trail at least $0.1$pc (assuming $1$m/s speed of escape)
along the system's trajectory in 100 Myrs, the time between disk
crossings.  The time for the solar system to cross a stream 
at a general angle will be longer than 50 years, the time 
to cross a bound microhalo (Diemand et al. 2005).  
Moving inside the inhomogeneous debris of 
the microhalos could cause
transient enhancements of the event rate in direct detection experiments. Furthermore clumps on polar orbits survive longer than planar orbits (cf.
Fig.~\ref{angus1}),  this makes the velocity distribution of the micro-streams non-Gaussian, 
and could imprint an interesting annual modulation on the
direct detection signal (Morgan et al. 2005; Freese et al. 2005). Fig.12 of Angus \& Zhao (2006) shows the morphology of a microstream: even after the microhalo is fully disrupted, the microstream is still a distinct entity with a negligable tangential size compared to its length. As a result, these filamentary streams may be the fate of a large proportion of the Milky Way's microhalos and the enhanced background from these structures may impose interesting features on direct detection experiments as well as having interesting annihilation flux properties.

The implications of dark microhalos and microstreams are rather different for the
case of indirect detection. Indirect measurements sample the
distribution of dark matter, through its annihilation rate, over large regions of the halo. 
Furthermore, dark
matter annihilation rates, and thus indirect detection signals,
are proportional not to the density of dark matter but to the dark matter density squared.
Substructure within the galactic halo thus may be capable of
boosting the dark matter annihilation rate and enhancing the
prospects of detecting dark matter indirectly (Taylor \& Silk 2003).

Techniques employed  for the indirect detection of dark matter
include gamma-ray, anti-matter and neutrino detectors (Silk \& Srednicki 1984). 
Gamma-rays annihilating in nearby and dense substructures, if present, 
could provide point sources potentially observable by gamma-ray telescopes (Pieri et al. 2005; Koushiappas et al. 2004). 
Anti-matter produced 
in dark matter annihilations are deflected by galactic magnetic fields and thus only the diffuse spectrum can be
studied. Neutrinos are not as useful for identifying
annihilations in the galactic halo, but instead are used to search
for dark matter particles annihilating in the core of the Sun. The
neutrino flux produced through dark matter matter annihilations in the Sun is tied to the local density
averaged over very long periods of time and therefore the
presence of substructure is of little importance.

Nearby microhalos, being tidally disrupted, are not
particularly bright point sources of gamma-rays. 
%%%Gamma-ray flux from a central cusp of size $r$, with a density that goes as %%%$r^{3-2\gamma}$ with $0 \le \gamma \le 1.2$ for the microhalos\cite{moore}. 
While Diemand et al. 2005 predict that the nearest intact microhalo  
will be much brighter than known dwarf spheroidals in gamma-rays, we find that the situation reverses
for a microhalo left with only a one percent bound remnant.  

Anti-matter fluxes produced through dark matter annihilations do not 
depend on the nature or location of individual dark substructures, 
but rather on the distribution of dark matter averaged over large
volumes. The value of the quantity $\left<\rho^2\right>/\left<\rho\right>^2$, 
averaged over the contributing volume (a few kiloparces for positrons or 
tens of kiloparsecs for anti-protons), determines the overall boost factor for
the anti-matter fluxes generated through dark matter annihilations. 

With a distribution of substructure of the form $dN/d\log M
\propto M^{-1}$, each decade of mass contributes almost equally
to the boost factor, so it is important to determine over
what range of masses substructures can survive.
Previous works (Helmi et al. 2002; Stoehr et al. 2003) emphasized the role of
largers substructures, with the mass of Draco dwarf spheroidal. 
Diemand et al. 2005 argue that the nearest Earth-mass halo is factor of a few
brighter than Draco.  We argue, however, that the nearest Earth-mass halo is likely
to be tidally dispersed, and hence fainter than Draco.
We have proposed (Zhao \& Silk 2005) that the nurseries of the first stars in the
universe are dense minihalos of $10^6\msun$ formed at $z \sim 22$.  Unlike the microhalos
such minihalos can survive the galactic tides until the present day.  The nearest minihalo is
only 1-2 kpc from the Sun, and is an order of magnitude brighter than
the combined annihilation from the Draco galaxy.

In light of the excess reported by the HEAT collaboration
(Barwick et al. 1997; Coutu et al. 1999; Baltz \& Edsjo 1999), substructure is of particular importance for dark matter searches
involving positrons.  To produce such a signal with thermally generated
neutralinos or other type of WIMPs, however, 
boost factors of $\sim$50 or higher are required (Baltz et al. 2005; Hooper \& Kribs 2004; Hooper \& Servant 2005). Although it
has been shown that more massive substructures do not naturally
provide such a large boost factor (Hooper et al. 2004), 
effects due to baryons, or the presence of intermediate mass black holes could perhaps make them brighter (Prada et al. 2004), although probably not enough to account for the HEAT excess. After
the effects of tidal disruption are considered, the same is found
to be true for Earth-mass microhalos. We agree with previous
studies (Berezinsky et al. 2003) which determine that positron boost
factors larger than $\sim$2-5 are unlikely. 
Given this conclusion,
another explanation for the excess observed by HEAT appears to be
required. Our understanding of
these issues will be dramatically improved with data from
the up-coming cosmic anti-matter experiments PAMELA and AMS-02
(Profumo \& Ullio 2004). 

\section{Discussion and Conclusions}

Using a semi-analytic estimate of the tidal 
mass-loss rate, together with numerical integration of a fair sample of microhalo orbits,
we find that 
most microhalos present in the solar neighborhood will have been heavily stripped by
stellar encounters, producing 'microstreams' of tidal debris. 
More generally, in environments with very low stellar densities such as the outer parts 
of the Galactic disk or in a Sextans-like dwarf galaxy, microhalos are only mildly heated. 
But in high-density environments such as the Galactic bulge or an M32-like elliptical 
($800$ to $0.05\msun\pc^{-3}$ in inner 1kpc of M32 (Mateo 1998)) microhalos are likely fully destroyed. In disk galaxies such as the Milky Way, the fraction of microhalos which are disrupted depends strongly on the orbital inclination and pericentre.
Microhalos on orbits coplanar with the disk are very quickly disrupted. 
Microhalos in the outer part 
of galaxy halos will encounter far fewer stars, however, and thus 
their annihilation rates will be less dramatically reduced. A possible consequence 
of this might be to detect little gamma-ray emission from the central 
part of external galaxies, where stars have disrupted most of the dark 
substructure, and more flux from the outer regions. A ring of gamma-rays 
surrounding a galaxy, if detected, would provide a strong confirmation 
of the existence of dark substructure. 

The tidal effects discussed in this paper and in 
several preprints appeared during the marathon referee process of this paper 
(Berezinsky, Dokuchaev, Eroshenko 2005, astro-ph/0511494, 
Green \& Goodwin 2006, astro-ph/0604142, 
Goerdt, Gnedin, Moore, Diemand, Stadel 2006, astro-ph/0608495,
Angus \& Zhao 2006, astro-ph/0608580) have important consequences for
the direct and indirect detection of dark matter.  In particular, the probability 
of the Earth intersecting a tidal stream at any given time is considerably larger 
than the probability of it being inside a microhalo's bound remnant, and thus tidal streams 
may potentially increase the reach of direct detection experiments. Indirect detection rates are not enhanced, however, as once stellar encounters are considered, Earth mass halos are not a significant contributor to the overall dark matter annihilation rate.

HSZ acknowledges support from PPARC Advanced Fellowship and Outstanding Oversea Young Scholarship from Chinese Academy of Science.  GWA acknowledges an overseas fieldwork grant from PPARC and hospitality from Beijing University.  JET acknowledges financial support from the NSF (grant AST-0307859)
and the DoE (contract DE-FG02-04ER41316).

%\vfill\eject

\vskip 1 cm

%\begin{figure}
%%\begin{center}
%\resizebox{10cm}{!}{
%\includegraphics{oldfig2.ps}
%\includegraphics{oldfig1a.ps}
%\includegraphics{oldfig1b.ps}}
%\caption{
%OLD figures (scaled to same size as new figure 1) for comparison ONLY, not 
%for publication.}
%%\label{starRZ}
%%\label{denv}
%%\end{center}
%\end{figure}

\end{document}